\documentclass[12pt]{article}
\usepackage{amsmath,amssymb,graphicx,bm,physics}
\usepackage{makecell}
\begin{document}

\title{Pseudo-Complex Gravity as a Geometric Resolution of the Black
  Hole Information Paradox}
	
\author{Fridolin Weber$^{1,2}$, Pter O. Hess$^{3,4}, Ceser A. Zen Vasconcellos^{5,6}$\\
{\small\it
$^1$Department of Physics, San Diego
  State University (SDSU), San Diego, 92182, USA}\\
{\small\it
$^2$Department of Physics, University of California at San
  Diego (UCSD),
La Jolla, CA 92093, USA}\\
{\small\it
$^3$Instituto de Ciencias Nucleares,
  UNAM, A.P. 70-543, 04510, Mexico}\\
{\small\it
$^4$Frankfurt Institute for Advanced
  Studies (FIAS), J.W. von Goethe Universit\"at, 60438 Frankfurt am
  Main, Germany}\\
{\small\it 
$^5$Instituto de Física,
  UFRGS,
	Porto Alegre
  91501-970, Brazil}\\
{\small\it
$^6$ICRANet,
65122 Pescara, Italy}
}

\maketitle
\abstract{
We investigate the black hole information paradox in the setting of
pseudo-complex gravity, a covariant geometric extension of general
relativity that introduces a minimal length scale by deforming the
spacetime manifold. In this framework, curvature invariants stay
finite, and the classical singularity is geometrically regularized via
a smooth core. We show that the correction term $B/(6 r^4)$ alters the
Schwarzschild metric, generating the regularized geometry above,
yielding a finite Hawking temperature, and inducing subleading
corrections to the Bekenstein–Hawking entropy.

Crucially, we demonstrate that the pseudo-complex geometric structure
obstructs a clean factorization of the Hilbert space into interior and
exterior regions, thereby removing the key assumption behind the
standard derivation of the paradox. This structural reinterpretation
of entanglement flow offers a new geometric route to unitarity
preservation and information recovery.

We examine the resulting effects on evaporation dynamics, entropy
flow, and thermodynamic behavior. Our predictions are compared with
those of generalized uncertainty principles (GUP), loop quantum
gravity (LQG), and island-based models, and are summarized in a
comparative table. Observable signatures—such as shifts in
quasi-normal mode frequencies and the appearance of gravitational wave
echoes from the regularized core—suggest that pseudo-complex gravity
is a testable, covariant approach to resolving the paradox without
invoking firewalls, holography, or exotic quantum states.
}

\section{Introduction}

The black hole information paradox remains one of the most enduring
and profound puzzles at the intersection of general relativity and
quantum mechanics. In classical general relativity, the evolution of a
gravitationally collapsing system leads to the formation of an event
horizon and, ultimately, a singularity. Hawking's semiclassical
derivation of black hole radiation~\cite{Hawking1975} compounded this
picture by showing that black holes evaporate thermally, with the
implication that information about the initial state is irretrievably
lost. This apparent violation of unitarity poses a deep conceptual
challenge to the foundations of quantum gravity.

Numerous proposals have been put forward to resolve the paradox,
including string-theoretic constructions such as
fuzzballs~\cite{Mathur2005}, the island formula and quantum extremal
surfaces~\cite{Penington2019, Almheiri2019}, and modifications of the
semiclassical framework via non-local effects or violations of
effective field theory. While each of these models has its own merits
and limitations, a common difficulty lies in establishing a robust,
covariant mechanism for regulating ultraviolet divergences near the
horizon and at the classical singularity.

Among the proposed resolutions, a common challenge is the absence of a
covariant, geometrically consistent mechanism for ultraviolet
regularization near the horizon and the classical
singularity. Pseudo-complex gravity, first proposed by Hess and
Greiner~\cite{HessGreiner2009,Hess2016,Hess2020}, addresses this by
extending the spacetime manifold through the promotion of coordinates
to pseudo-complex variables of the form \( X^\mu = x^\mu + I y^\mu \),
where \( I^2 = +1 \).\footnote{Pseudo-complex refers to a specific
algebraic extension with \( I^2 = +1 \), distinct from complexified
coordinates where \( i^2 = -1 \).} This structure introduces a
built-in minimal length scale, implemented covariantly via a
constraint on the pseudo-imaginary components \( y^\mu \), which can
be interpreted as encoding a maximal acceleration. The theory retains
the smooth manifold structure and diffeomorphism invariance of general
relativity, while embedding an ultraviolet cutoff directly into the
geometry.

In this work, we explore the implications of pc-Gravity for the black
hole information paradox. We begin by analyzing the structure of
Schwarzschild black holes in the pseudo-complex framework, showing
that the singularity is removed and the horizon geometry is modified.
(In particular, we compute the Kretschmann scalar and demonstrate, in
Eq.~(\ref{eq:K_pc}), that its divergence is softened by a geometric
correction.)  We compute corrections to the Bekenstein--Hawking
entropy that emerge from this regularization and discuss their
consistency with known quantum gravity models. We then examine how
pseudo-complex geometry influences the causal structure, entropy
counting, and potential mechanisms for unitarity restoration.

Although our analysis is primarily theoretical, we also comment on the
observational implications of the pc-Gravity framework, including
modifications to black hole shadows, gravitational wave ringdowns, and
potential echo signals. These features suggest that pseudo-complex
geometry  offers not only a novel perspective on the information
paradox, but also  a potentially testable deviation from
classical general relativity in strong-gravity regimes.

In addition to addressing foundational questions, pseudo-complex
gravity yields distinctive observational predictions.  These
predictions are increasingly relevant as gravitational wave astronomy
and black hole imaging reach precision levels capable of probing
near-horizon structure.  The predictions include subleading corrections to
quasi-normal mode frequencies, Planck-scale deviations from thermal
evaporation, and late-time gravitational wave echoes. We compare these
features with those predicted by other leading frameworks—such as the
generalized uncertainty principle (GUP) and loop quantum gravity
(LQG)—and assess their prospects for empirical constraint in current
and future observational programs.

The paper is structured as follows. In Section~\ref{sec:pc-gravity},
we review the mathematical framework of pseudo-complex gravity,
including its algebraic foundation, the constraint enforcing a maximal
acceleration, and an explicit derivation illustrating curvature
regularization near the Schwarzschild singularity. In
Section~\ref{sec:entropy}, we compute corrections to the black hole
entropy and Hawking temperature induced by the pseudo-complex
regulator, and demonstrate thermodynamic consistency using both
geometric and first-law methods. Section~\ref{sec:unitarity} discusses
how these modifications alter the evaporation process, entropy flow,
and causal structure, with an emphasis on implications for
unitarity. We also examine the Hilbert space structure in the
pseudo-complex context and propose a reinterpretation of entanglement
entropy.

In Section~\ref{sec:comparison}, we compare pseudo-complex gravity
with other major approaches to the black hole information paradox,
including firewalls, fuzzballs, the island rule, and GUP- or
LQG-inspired models. Section~\ref{sec:observables} presents potential
observational signatures, including shifts in quasi-normal mode (QNM)
frequencies, gravitational wave echoes, and early-universe
imprints. Section~\ref{sec:qnm-comparison} provides a comparative
analysis of the QNM spectrum across pc-Gravity, GUP, and LQG
frameworks, and estimates the magnitude of deviations expected near
the Planck scale. We conclude in Section~\ref{sec:conclusion} with a
summary of results and a discussion of promising directions for future
work, including extensions to rotating black holes and quantum
formulations of pseudo-complex gravity.

Technical derivations supporting the scalar two-point function,
corrected thermodynamic quantities, and a proposed quantization scheme
over pseudo-complex manifolds appear in
\ref{sec:appendix-a}, \ref{sec:appendix-b},
and~\ref{sec:appendix-c}, respectively.

Throughout this paper, we work in natural units where \( \hbar = c = G
= k_B = 1 \). In this system, all quantities are expressed in powers
of the Planck units: length \( \ell_P \), mass \( m_P \), time \( t_P
\), and temperature \( T_P \). Entropy is dimensionless and measured
in units of Boltzmann’s constant, while black hole mass and
temperature are typically given in Planck units unless otherwise
stated.

\medskip

\section{Overview of Pseudo-Complex Gravity}\label{sec:pc-gravity}

Pseudo-complex gravity extends the geometric framework of general
relativity by promoting spacetime coordinates to elements of a
pseudo-complex algebra. A point in spacetime is described by
\cite{HessGreiner2009,Hess2016,Hess2020}
\begin{equation}
  X^\mu = x^\mu + I y^\mu, \quad I^2 = +1,
  \label{eq:X.mu}
\end{equation}
where \( x^\mu \in \mathbb{R}^4 \) are the usual spacetime coordinates
and \( y^\mu \in \mathbb{R}^4 \) are pseudo-imaginary components
encoding internal degrees of freedom associated with a minimal length
scale.

The algebra of pseudo-complex numbers \( \mathbb{P} \) is commutative,
associative, and contains zero divisors. This structure admits an
idempotent decomposition, given in Eq.~(\ref{eq:e.pm}),
\begin{equation}
  e_\pm = \frac{1}{2}(1 \pm I), \quad e_\pm^2 = e_\pm, \quad e_+ e_- = 0,
  \label{eq:e.pm}
\end{equation}
so that any pseudo-complex quantity can be written as
\begin{equation}
  X^\mu = X_+^\mu e_+ + X_-^\mu e_- ,
  \label{eq:X.mu.2}
\end{equation}
with \( X^\mu_\pm \in \mathbb{R}^4 \). In this basis, the
pseudo-complex spacetime manifold can be viewed as a direct sum of two
real manifolds governed by different geometric sectors.

The spacetime metric is likewise extended to a pseudo-complex-valued tensor field
\begin{equation}
G_{\mu\nu}(X) = g_{\mu\nu}(x, y) + I h_{\mu\nu}(x, y),
\end{equation}
leading to the line element
\begin{eqnarray}
dS^2 &=& G_{\mu\nu}(X) \, dX^\mu dX^\nu \nonumber \\
     &=& g_{\mu\nu}(dx^\mu dx^\nu + dy^\mu dy^\nu) + 2 h_{\mu\nu} dx^\mu dy^\nu \nonumber \\
     &&+ I h_{\mu\nu}(dx^\mu dx^\nu + dy^\nu dy^\mu) + 2 I g_{\mu\nu} dx^\mu dy^\nu .
\end{eqnarray}
This decomposition reflects the full coordinate structure as expressed
in Eq.~(\ref{eq:X.mu.2}).

Since the metric component \( g_{tt} \) is assumed to be real in the
present treatment, and because both \( g_{\mu\nu} \) and \( h_{\mu\nu}
\) are symmetric, we adopt the simplification \( h_{\mu\nu} = 0 \), as
motivated in earlier work~\cite{Hess2016}. This assumption
retains the pseudo-complex structure in the kinetic sector without
introducing explicit imaginary parts into the metric.

To maintain physical viability, pc-Gravity imposes a constraint on the
pseudo-imaginary four-velocity \( \dot{y}^\mu = dy^\mu / d\tau \) that
enforces a maximal acceleration,
\begin{equation}
g_{\mu\nu} \dot{y}^\mu \dot{y}^\nu = \frac{1}{a_{\rm{max}}^2}, \quad
\rm{with} \quad a_{\rm{max}} \sim \frac{c^2}{\ell},
\label{eq:g.munu.6}
\end{equation}
where \( \ell \) is the minimal length scale.  The maximal
acceleration constraint, shown in Eq.~(\ref{eq:g.munu.6}), regulates
short-distance behavior by suppressing arbitrarily large proper
accelerations.  It is analogous to the speed-of-light limit in special
relativity.

Physically, the \( y^\mu \) directions can be interpreted as encoding
geometric deformations that act as internal regulators, smoothing
curvature singularities and enforcing geodesic
completeness. Importantly, the pseudo-complex extension does not
introduce discretization or break general covariance. Instead, it
embeds an ultraviolet cutoff directly into the geometric structure of
spacetime.

\subsection*{Curvature Regularization: An Explicit Example}

To illustrate the regularizing power of pc-Gravity, consider the
modified Schwarzschild-like metric
\begin{eqnarray}
  ds^2 =&& \left(1 - \frac{2GM}{c^2 r} + \frac{B}{6 r^4} \right) c^2 dt^2
 \nonumber \\
  &&-
\left(1 - \frac{2GM}{c^2 r} + \frac{B}{6 r^4} \right)^{-1} dr^2 - r^2
d\Omega^2,
 \label{eq:ds2Schw} 
\end{eqnarray}
where \( B \propto \ell^4 \), consistent with the dimensional scaling
of the correction term.  This form of the correction term \( B/(6r^4)
\) is motivated by previous applications of pseudo-complex gravity to
regular black hole metrics and is compatible with post-Newtonian
constraints at solar system scales \cite{Hess2016}.
%The correction term \( B/(6r^4) \) is chosen to remain consistent with
%post-Newtonian constraints while capturing near-horizon regularization
%in pseudo-complex gravity~\cite{Hess2016}.
For standard Schwarzschild geometry, the
Kretschmann scalar diverges as
\begin{equation}
\mathcal{K}_{\rm{GR}} = R_{\mu\nu\rho\sigma} R^{\mu\nu\rho\sigma} =
\frac{48 G^2 M^2}{c^4 r^6}.
\end{equation}
In the pseudo-complex corrected case, this becomes
\begin{equation}
K_{\rm{pc}} = \frac{48G^2M^2}{c^4 r^6} \left(1 - \frac{\alpha
  B}{r^4} + \mathcal{O}(B^2)\right) .
  \label{eq:K_pc}
\end{equation}
As shown in Eq.~(\ref{eq:K_pc}), the divergence in the Kretschmann
scalar is softened by the pseudo-complex correction. As \( r \to 0 \),
the \( B/r^4 \) term dominates (with \( B \sim \ell^4 \)) and
suppresses the divergence. If \( B = \ell^4 > 0 \), the scalar
curvature invariants remain finite or at least less singular than in
general relativity (see Eq.~(\ref{eq:K_pc})).

This mechanism is conceptually related to earlier nonsingular black
hole models, such as the Hayward solution~\cite{Hayward2006}; to
horizonless proposals like black stars~\cite{Barcelo2009}, which
replace classical horizons with smooth, compact cores supported by
quantum effects; and to Frolov’s closed-apparent-horizon
scenario~\cite{Frolov2014Nonsingular}. In contrast, pc-Gravity
achieves regularization through a covariant geometric deformation
rather than through an effective matter source. It is also comparable
to quantum-corrected evaporation models in loop quantum gravity, such
as that proposed by Ashtekar and Bojowald~\cite{AshtekarBojowald2005}.

\subsection*{Classical Limit and Strong-Field Corrections}

In the limit \( y^\mu \to 0 \), or equivalently \( B \to 0 \), the
pseudo-complex theory reduces smoothly to classical general
relativity. In strong-field regimes—such as those near black hole
horizons or in early-universe cosmology—the pseudo-complex
contributions become significant, potentially altering causal
structure, horizon dynamics, and the thermodynamics of spacetime.

In the following sections, we investigate how these geometric
modifications influence black hole entropy, radiation, and the
information flow across horizons, with a view toward a resolution of
the black hole information paradox.

\begin{figure}[htb]
    \centering
    \includegraphics[width=0.5\linewidth]{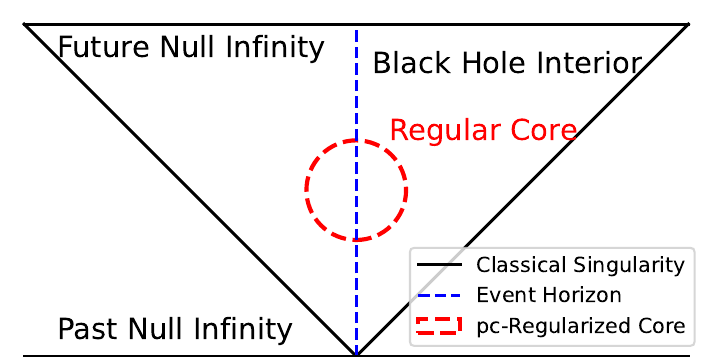}
    \caption{Penrose-like diagram comparing the classical
      Schwarzschild causal structure with the pseudo-complex gravity
      regularized core. In standard GR, null geodesics terminate at a
      curvature singularity located at \( r = 0 \), shown here as a
      black diagonal boundary. In pc-Gravity, this singularity is
      replaced by a smooth, timelike region (red dashed circle) that
      reflects the presence of a minimal length scale and a maximal
      proper acceleration. The event horizon (blue dashed line)
      remains a causal boundary but now encloses a regular core
      through which geodesics may be continued. This modification
      supports geodesic completeness and allows for causal contact
      between interior and exterior regions, potentially enabling
      unitary information flow during evaporation.  }
    \label{fig:pc_penrose_diagram}
\end{figure}

Figure~\ref{fig:pc_penrose_diagram} schematically illustrates how the
causal structure of Schwarzschild spacetime is modified in
pseudo-complex gravity. In the classical picture, null geodesics
terminate at a spacelike singularity, signaling the breakdown of
predictability. In contrast, the pseudo-complex extension replaces the
singularity with a regular core, represented by a timelike surface,
which imposes a lower bound on proper acceleration and curvature. This
regularized region allows geodesics to continue smoothly through the
would-be singularity, potentially enabling causal contact between the
interior and the external universe. Such modifications lay the
geometric groundwork for avoiding information loss during black hole
evaporation.

\section{Entropy Corrections and Finite State Counting}\label{sec:entropy}

A key consequence of the pseudo-complex extension of gravity is the
introduction of a covariant ultraviolet regulator that modifies
the area-entropy relation for black holes. In standard general
relativity, the Bekenstein--Hawking entropy of a Schwarzschild black
hole is given by
\begin{equation}
S_{\rm{BH}} = \frac{k_B c^3}{4 G \hbar} A = \frac{k_B A}{4 \ell_P^2},
\label{eq:S_BH}
\end{equation}
where \( A = 4\pi r_s^2 \) is the area of the event horizon and \(
\ell_P = \sqrt{\hbar G / c^3} \) is the Planck length.

In pseudo-complex gravity, the Schwarzschild metric is modified by a
term \( B/(6 r^4) \), as seen in Eq.~(\ref{eq:ds2Schw}), resulting in
a shifted event horizon:
\begin{equation}
r_+ \approx \frac{2GM}{c^2} - \frac{B c^2}{24 G^2 M^3},
\end{equation}
and a corrected area
\begin{equation}
A = 4\pi r_+^2 \approx 16 \pi \frac{G^2 M^2}{c^4} \left(1 - \frac{B
  c^6}{12 G^2 M^4} \right).
\end{equation}
In the limit \( B \to 0 \), the pseudo-complex corrected horizon
radius \( r_+ \) reduces to the classical Schwarzschild radius \( r_s
= 2GM/c^2 \), so that \( r_+ \approx r_s \) up to small corrections.
This recovers the standard Bekenstein–Hawking entropy expression given
in Eq.~(\ref{eq:S_BH}), which forms the baseline for subsequent
pseudo-complex corrections.

\begin{figure}[tb]
    \centering
    \includegraphics[width=0.6\linewidth]{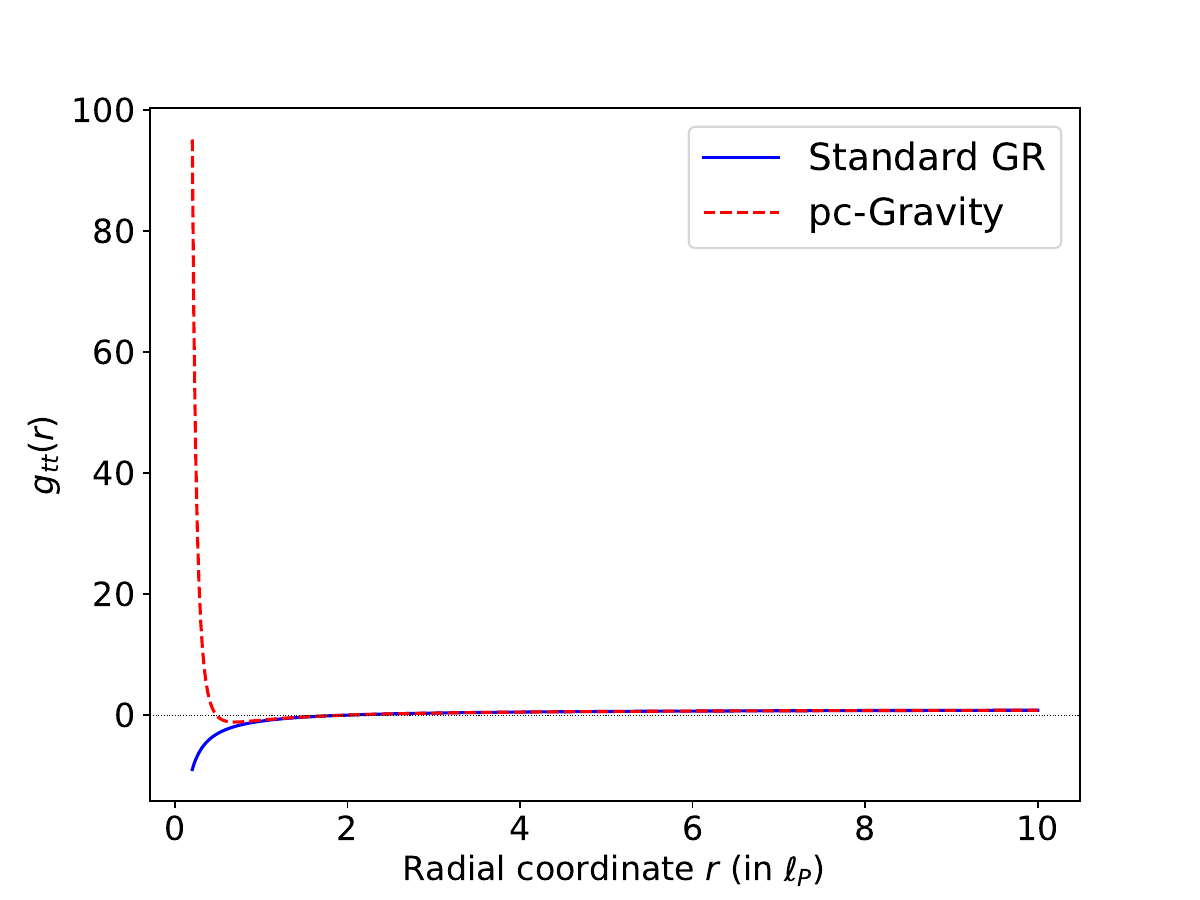}
    \caption{ Comparison of the time-time component \( g_{tt}(r) \) of
      the Schwarzschild metric in standard general relativity (blue)
      and in pseudo-complex gravity (red dashed). In classical GR, \(
      g_{tt}(r) \) becomes singular at \( r \to 0 \) and crosses zero
      at the Schwarzschild radius \( r_s = 2GM/c^2 \). In pc-Gravity,
      the inclusion of a geometric correction term \( B/(6 r^4) \) softens
      the near-singularity behavior, regularizing the curvature and
      shifting the horizon inward. This demonstrates how
      pseudo-complex geometry eliminates the central singularity while
      preserving asymptotic Schwarzschild behavior, enabling geodesic
      completeness and possibly altering causal structure near the
      core.  }
    \label{fig:pc_metric_regularization}
\end{figure}

Figure~\ref{fig:pc_metric_regularization} illustrates the behavior of
the time-time component \( g_{tt}(r) \) in both the standard
Schwarzschild solution and the pseudo-complex corrected geometry. In
classical general relativity, \( g_{tt} \) drops sharply and diverges
as \( r \to 0 \), reflecting the presence of a curvature
singularity. In contrast, the pseudo-complex correction introduces a
\( B/(6r^4) \) term that becomes dominant at small \( r \), softening the
divergence and rendering the metric component finite throughout the
entire spacetime manifold. The event horizon is shifted slightly
inward, as seen from the change in the root of \( g_{tt}(r) = 0
\). This regularized behavior supports the interpretation of
pc-Gravity as a covariant framework in which geodesics are extendable
through the core, potentially preserving information and altering the
standard causal structure associated with black holes.

To reinforce this correction with thermodynamic consistency, consider
the first law of black hole thermodynamics,
\begin{equation}
  dM = T_H \, dS,
  \label{eq:13}  
\end{equation}
where \( T_H \) is the Hawking temperature. The surface gravity for
the modified metric is given by
\begin{equation}
  \kappa = \frac{GM}{c^2 r_+^2} - \frac{B}{3 r_+^5},
  \label{eq:14}
\end{equation}
leading to a corrected temperature,
\begin{equation}
  T_H^{\rm pc} = \frac{\hbar \kappa}{2\pi k_B} \approx \frac{\hbar
    c^3}{8\pi G M k_B} \left(1 - \frac{5 B c^4}{24 G^2 M^4} \right).
  \label{eq:T_H_pc}
\end{equation}

Using Eq.~(\ref{eq:13}), together with the corrected surface gravity
from Eq.~(\ref{eq:14}) and the modified temperature in
Eq.~(\ref{eq:T_H_pc}), we obtain for the pseudo-complex corrected entropy
\begin{equation}
  S_{\rm pc} = \int \frac{dM}{T_H^{\rm pc}} \approx \frac{4\pi k_B G
    M^2}{\hbar c} \left(1 - \frac{B c^4}{12 G^2 M^4} \right).
  \label{eq:S_pc.a}
\end{equation}
This matches the corrected entropy obtained from the area expression:
\begin{equation}
  S_{\rm pc} = S_{\rm BH} \left(1 - \frac{B c^4}{12 G^2 M^4} \right).
  \label{eq:S_pc.b}
\end{equation}
The result given in Eq.~(\ref{eq:S_pc.b}) shows that entropy growth
slows at small $M$.  Similar entropy corrections have also been
derived in loop quantum gravity, where quantum geometric effects
modify the semiclassical area law~\cite{AshtekarBojowald2005}.

\begin{figure}[tb]
    \centering
    \includegraphics[width=\linewidth]{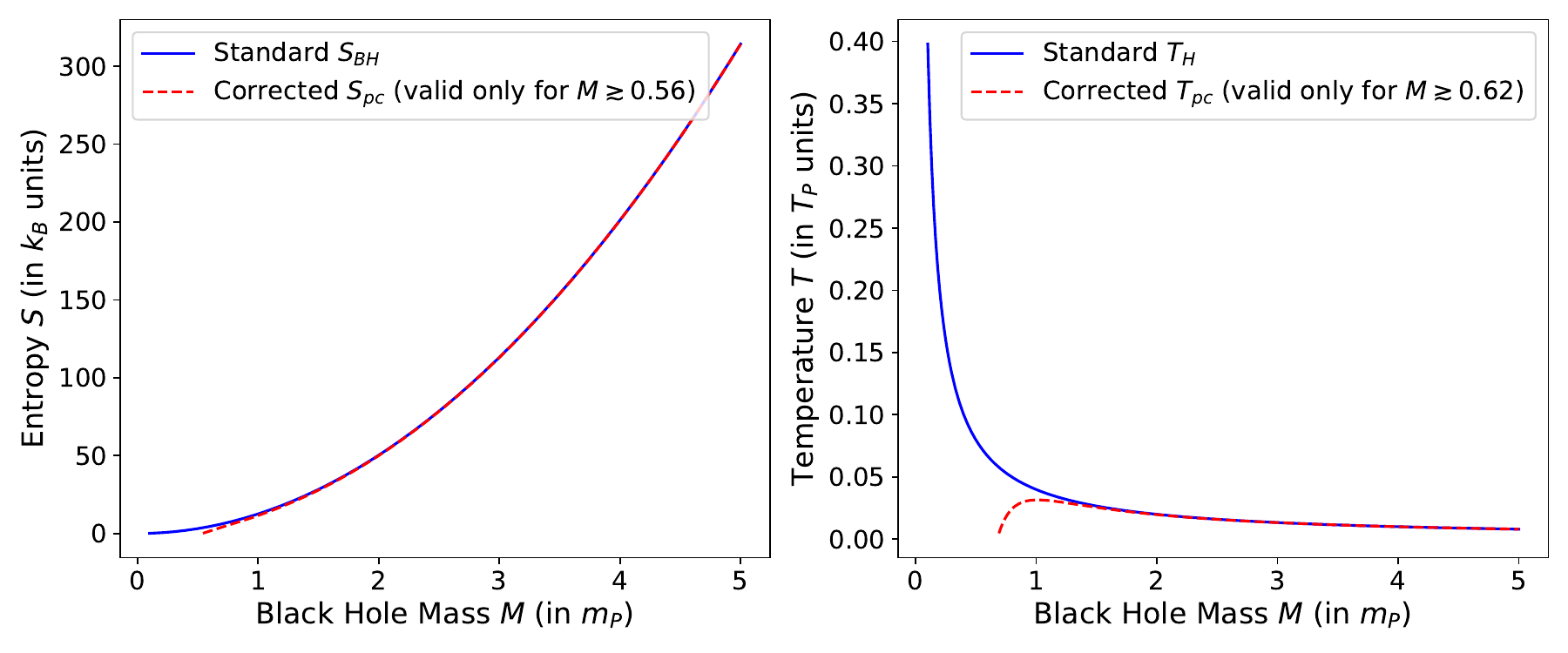}
    \caption{Left: Comparison of standard Bekenstein--Hawking entropy
      \( S_{\mathrm{BH}} \) and pseudo-complex corrected entropy \(
      S_{\mathrm{pc}} \) (dimensionless, in units of \( k_B \)) as a
      function of black hole mass \( M \) (in Planck units). The
      correction introduces a subleading \( 1/M^4 \) term that
      suppresses entropy at small masses, suggesting a finite density
      of microstates. Right: Hawking temperature \( T \) (in units of
      the Planck temperature) vs. mass for the standard case and
      pseudo-complex corrected expression. The correction lowers the
      temperature in the ultraviolet regime, implying slower
      evaporation near the Planck scale. For this figure, we set \( B
      = \ell^4 = 1 \), corresponding to a minimal length \( \ell \)
      equal to the Planck length in natural units.}
    \label{fig:pc_entropy_temperature}
\end{figure}

Figure~\ref{fig:pc_entropy_temperature} illustrates the impact of
pseudo-complex corrections on black hole entropy and Hawking
temperature, as shown in Eq.~(\ref{eq:S_pc.b}) and plotted in the left
panel of Fig.~\ref{fig:pc_entropy_temperature}.  As shown in the left
panel, the corrected entropy \( S_{\mathrm{pc}} \) grows more slowly
than the standard Bekenstein--Hawking entropy \( S_{\mathrm{BH}} \)
for small black hole masses. This deviation becomes significant near
the Planck scale and
reflects the suppression of trans-Planckian degrees of freedom due to
the pseudo-complex geometric regulator, a feature also present in
quantum-corrected black hole evaporation models in loop quantum
gravity~\cite{AshtekarBojowald2005}.
The right panel displays the
corresponding Hawking temperature profiles. In the pseudo-complex
case, the surface gravity is reduced, leading to a lower temperature
for a given mass compared to the standard result. These effects
suggest that evaporation slows at high curvature, and that a complete
evaporation process may avoid singularity formation altogether.  This
scenario aligns with the broader class of regular black hole models
that allow for nonsingular endpoints, as first explored
in~\cite{Hayward2006}.
This is consistent with expectations from non-singular black hole
models that feature modified core geometries and remnant
formation~\cite{Hayward2006}.
The
joint behavior of the entropy and temperature supports the idea that
pseudo-complex gravity enforces a geometric form of ultraviolet
completeness while remaining consistent with classical thermodynamic
principles at large scales.

This subleading correction is negative and scales as \( 1/M^4 \),
becoming significant for black holes near the Planck mass. Such
behavior suggests a suppression of entropy in the UV, implying a
finite density of microstates. This is consistent with expectations
from generalized uncertainty principles, loop quantum gravity, and
non-commutative geometry, yet pc-Gravity achieves it through a purely
geometric and covariant mechanism without modifying commutation
relations or invoking discrete spectra.  The suppression of entropy at
low mass, shown in Eq.~(\ref{eq:S_pc.b}), indicates a finite number of
microstates and a possible endpoint to evaporation.

Furthermore, the entropy correction signals a deviation from strict
thermality. As we show next, this opens a window for
unitarity-preserving radiation and quantum information flow.

\section{Toward Unitarity: Modified Radiation and Information Flow}\label{sec:unitarity}

The semiclassical account of black hole evaporation predicts that the
emitted radiation is perfectly thermal. If this is strictly true, the
evaporation process would lead to a final state that is fundamentally
mixed, in direct conflict with the unitary evolution required by
quantum mechanics. This conflict lies at the heart of the black hole
information paradox~\cite{Hawking1975}.

The framework of pseudo-complex gravity offers a fresh approach to
this issue by modifying the underlying geometry of spacetime in a
covariant way. The presence of a regular, nonsingular core and a
built-in ultraviolet cutoff alters the assumptions that underlie
Hawking’s original calculation. The event horizon in this context no
longer represents a sharply defined causal boundary, but instead
admits a modified structure that allows for continued causal contact
across regions that would otherwise be classically disconnected. The
curvature near the core is smoothed, enabling the extension of
geodesics through what would otherwise be a singularity. Furthermore,
the entropy–area relation receives corrections that suppress the
degeneracy of radiation states at small scales, indicating a departure
from strict thermality and hinting at the possibility of information
recovery during evaporation.

The corrected Hawking temperature leads to a modified evaporation
profile. Since \( T_H^{\rm{pc}} < T_H \) for a given mass, the rate
of mass loss is suppressed,
\begin{equation}
\frac{dM}{dt} \propto - A (T_H^{\rm{pc}})^4,
\end{equation}
implying longer lifetimes for small black holes and possibly the
formation of remnants ~\cite{Giddings1992}.

Additionally, the reduced entropy at low mass suggests that the
information content of the black hole may be encoded in a finite set
of degrees of freedom. As radiation proceeds, deviations from strict
thermality—arising from the pseudo-complex structure—could permit the
gradual leakage of information, in a manner consistent with the Page
curve.

Several structural features of pseudo-complex gravity contribute to
this possibility. The deformation of the radiation spectrum, induced
by modifications to the greybody factors and surface gravity, alters
the spectral profile in a way that departs from perfect thermality. At
the same time, the presence of a smooth, regular core modifies the
causal structure, enabling potential signal reflection or information
transfer that circumvents the singularity and preserves
continuity.
This behavior resembles black star models~\cite{Barcelo2009}, in which
the absence of a true horizon allows for reflective surfaces or
information re-emergence during evaporation.
Perhaps most significantly, the underlying algebraic
structure of pc-Gravity implies a fundamentally non-factorizable
Hilbert space, in which the standard inside/outside decomposition of
semiclassical gravity may no longer apply. These features collectively
offer a covariant, geometric pathway toward resolving the black hole
information paradox, without resorting to firewalls, non-locality, or
new dynamical fields. A similar goal is achieved in loop quantum
gravity through effective dynamics that preserve unitarity and
eliminate the singularity~\cite{AshtekarBojowald2005}.

Future work should focus on developing a detailed model of the full
evaporation process within this framework, and on exploring the
implications of pseudo-complex geometry for entanglement entropy, the
Page curve, and quantum extremal surfaces.

\subsection*{Hilbert Space Structure and Nonlocal Entanglement}

A central feature of the black hole information paradox is the
assumption that the Hilbert space describing quantum gravity can be
cleanly factorized into interior and exterior degrees of freedom, as
expressed in Eq.~(\ref{eq:H.19}).
 This
factorization is essential to tracing out the interior and
interpreting the radiation as thermal. However, this assumption may
not hold in pseudo-complex gravity.

The extended pseudo-complex structure introduces an additional set of
internal geometric degrees of freedom, governed by the
pseudo-imaginary coordinates \( y^\mu \), which are constrained by a
maximal acceleration condition. These components modify the structure
of the spacetime manifold itself, potentially entangling the UV
(short-distance) and IR (long-distance) degrees of freedom in a
nontrivial, nonlocal way—even though the spacetime remains covariant
and smooth.

This suggests that the full Hilbert space in pc-Gravity may not admit
a clean tensor product decomposition of the form
\begin{equation}
  \mathcal{H} = \mathcal{H}_{\rm{int}} \otimes \mathcal{H}_{\rm{ext}},
  \label{eq:H.19}
\end{equation}
as typically assumed in semiclassical gravity.  While a full
  quantum formulation of pseudo-complex gravity is not yet
  established, the algebraic structure suggests a decomposition of the
  Hilbert space into pseudo-complex sectors with geometric
  entanglement between them. Instead, the presence of nontrivial
idempotent sectors \( e_\pm \) may imply a decomposition into
overlapping or braided subspaces,
\begin{equation}
\mathcal{H}_{\rm{pc}} = \mathcal{H}_+ \oplus \mathcal{H}_-,
\end{equation}
with nontrivial correlations between them that are geometrically
encoded.

As a consequence, the entanglement entropy of the radiation may not
arise purely from tracing out inaccessible interior modes. Instead,
the pseudo-complex structure could preserve information across the
entire spacetime in a delocalized fashion. This reinterpretation of
entanglement flow aligns with ideas from modular flow and quantum
error correction in holography~\cite{Harlow2017}, though it arises
here from the intrinsic geometry rather than duality arguments.

Further work is needed to characterize the algebra of observables in
this extended framework and to determine whether pseudo-complex
corrections can reproduce known features such as the Page curve or
quantum extremal surfaces. Nevertheless, the built-in geometric
correlation structure offers a promising avenue for rethinking quantum
information flow in gravitational systems. A detailed dynamical
calculation of the Page curve in the pc-Gravity framework is an
important direction for future work.  A possible quantization scheme
over pseudo-complex manifolds is outlined in \ref{sec:appendix-c}.

\section{Comparison with Other Resolutions}\label{sec:comparison}

The black hole information paradox has inspired a broad spectrum of
theoretical proposals, each attempting to reconcile quantum mechanics
with the gravitational physics of black hole
evaporation. Pseudo-complex gravity adds to this landscape a
covariant, geometrically motivated framework that regularizes
curvature singularities and modifies horizon-scale physics without
introducing non-locality, external degrees of freedom, or discrete
spacetime structure.

\subsection{Firewalls and Fuzzballs}

The firewall hypothesis~\cite{Almheiri2013} posits a breakdown of
semiclassical physics at the event horizon, replacing the smooth
geometry with a high-energy "firewall" that destroys infalling
information. This proposal preserves unitarity but violates the
equivalence principle, challenging the foundations of general
relativity.

By contrast, the fuzzball proposal~\cite{Mathur2005} from string
theory replaces black holes with horizonless but complicated
microstate geometries. Each fuzzball corresponds to a particular state
of the black hole and emits non-thermal radiation from its
surface. While the fuzzball program has had success in certain
supersymmetric settings, its generalization to realistic, non-extremal
black holes remains incomplete.

pc-Gravity avoids the need for either firewalls or exotic microstate
geometries by modifying the horizon and singularity structure in a
smooth and covariant manner. The black hole retains a horizon-like
structure, but the absence of a singularity and the introduction of a
minimal length allow for potentially unitary evolution without
violating general relativity in the classical regime. The appearance of a minimal length scale is a recurring feature in quantum gravity scenarios \cite{Garay1995}.

\subsection{Quantum Extremal Surfaces and Islands}

Recent developments in holography have led to the proposal of the
"island rule"~\cite{Penington2019, Almheiri2019}, which computes the
entropy of Hawking radiation using quantum extremal surfaces that can
include portions of the black hole interior. This approach has
successfully reproduced the Page curve in several toy models and is
consistent with unitarity.

pc-Gravity may provide a geometric realization of similar behavior
without relying on a dual holographic theory. Because the
pseudo-complex structure modifies the causal and curvature structure
of spacetime, it may naturally give rise to effective quantum extremal
surfaces or entanglement wedge reconstructions in a covariant
setting. A detailed comparison remains to be developed, but the
mechanisms are complementary, holographic islands operate within
semiclassical gravity augmented by entanglement entropy, while
pc-Gravity modifies the spacetime geometry at a fundamental level.

\subsection{Non-Commutative Geometry, GUP, and Loop Gravity}

Approaches such as the generalized uncertainty
principle~\cite{Kempf1995}, non-commutative
geometry~\cite{Ashtekar1998,Nicolini2006}, and loop quantum
gravity~\cite{RovelliVidotto2014} all introduce minimal
length scales or fundamental discreteness in spacetime. These
frameworks predict quantum corrections to black hole entropy and
horizon structure, often yielding subleading terms that scale as
\(1/M^4\) or include logarithmic corrections. Such modifications are
typically interpreted as signatures of an underlying UV-complete
theory that resolves classical singularities and alters the structure
of spacetime near the horizon. For example, loop quantum gravity
implements singularity resolution through quantum geometry and has
been shown to yield unitary evaporation~\cite{AshtekarBojowald2005}.

Pseudo-complex gravity shares many of the qualitative features of
these approaches—most notably, the suppression of trans-Planckian
modes and the finiteness of black hole entropy—but implements them
through a purely geometric mechanism. It does not require modifying
the fundamental commutation relations, discretizing spacetime, or
invoking specific quantum gravity states, as is common in
non-commutative geometry or DSR-type
frameworks~\cite{Amelino-Camelia2001}. Unlike DSR, which introduces a
minimal length through deformation of Lorentz symmetry, pc-Gravity
realizes this scale via an extension of the geometric structure while
preserving classical diffeomorphism invariance. The pseudo-complex
formalism thus enables continuous, covariant metrics to encode
ultraviolet corrections to classical gravity in a natural and
algebraically tractable manner.

An important direction for further clarification is to examine whether
pseudo-complex gravity can be embedded into a low-energy effective
field theory (EFT) framework. Since pc-Gravity introduces a geometric
ultraviolet cutoff via a covariant deformation of the metric, it is
natural to ask whether its long-wavelength behavior can be captured by
a series of higher-curvature corrections to the Einstein–Hilbert
action.

In standard EFT language, one considers an action of the form
\begin{eqnarray}
S_{\rm EFT} &=& \frac{1}{16\pi G} \int d^4x \sqrt{-g} \Bigl( R +
\alpha_1 R^2 + \alpha_2 R_{\mu\nu} R^{\mu\nu}  \nonumber \\
&& \hspace{4em} + \alpha_3 R_{\mu\nu\rho\sigma} R^{\mu\nu\rho\sigma} + \cdots \Bigr),
\label{eq:21}
\end{eqnarray}
This is the curvature-expanded effective field theory to which the
pseudo-complex correction can be matched. The coefficients \( \alpha_i
\) have dimension \( \rm{(length)}^2 \) and are suppressed by the
scale of new physics. In pc-Gravity, the correction to the
Schwarzschild metric appears as a term of the form \( B/(6r^4) \) in
\( g_{tt}(r) \), which suggests an effective contribution that becomes
important at short distances. These contributions can be interpreted
in terms of the EFT action shown in Eq.~(\ref{eq:21}), where the
coefficients \( \alpha_i \) encode the strength of higher-curvature
corrections arising from the pseudo-complex deformation.

%This behavior qualitatively resembles that induced by certain quadratic curvature invariants in the near-horizon regime.

To make this correspondence precise, one could expand the
pseudo-complex corrected metric in powers of the small parameter \( B
\sim \ell^4 \), and identify the effective action that reproduces the
same leading-order corrections to curvature invariants such as the
Kretschmann scalar. This procedure would permit extraction of matching
conditions between pc-Gravity and a curvature-expanded EFT. For
example, the regularization of the \(1/r^6\) divergence in the
Kretschmann scalar could be mimicked by a specific combination of \(
R^2 \) and \( R_{\mu\nu\rho\sigma} R^{\mu\nu\rho\sigma} \) terms, with
the coefficients \( \alpha_i \) scaling as \( \ell^2 \).

Such a matching would clarify the relationship between pseudo-complex
geometry and higher-derivative corrections found in string theory,
asymptotically safe gravity, and loop quantum gravity. Crucially,
because the pseudo-complex extension preserves full covariance, it is
plausible that the EFT matching could be carried out
systematically. This would also shed light on whether the
regularization mechanism in pc-Gravity leads to additional propagating
degrees of freedom in the linearized theory, or whether it corresponds
purely to higher-derivative corrections without introducing new
physical modes.

A complete matching analysis—based on metric perturbation theory and
curvature functional expansions—remains a promising direction for
future research, and may serve as a bridge between the geometric
regularization mechanism of pc-Gravity and the EFT-based approach to
quantum gravity phenomenology.

A side-by-side comparison of structural and phenomenological features
of these models is presented in Table~\ref{tab:comparison}.

\begin{table}[htb]
\centering
\small
\caption{Comparison of key structural features and phenomenological
  predictions among pseudo-complex gravity (pc-Gravity), the
  generalized uncertainty principle (GUP), loop quantum gravity (LQG),
  and island-based models. Distinctions are evident in the origin of
  ultraviolet regularization, entropy corrections, and observability
  of quantum gravitational effects.}
\vspace{0.5em}
\renewcommand{\arraystretch}{1.3}
\resizebox{0.95\textwidth}{!}{%
\begin{tabular}{|l|c|c|c|c|}
\hline
\textbf{Feature} & \textbf{pc-Gravity} & \textbf{GUP} & \textbf{LQG} & \textbf{Islands} \\
\hline
UV cutoff origin & \makecell{Geometric \\ (pc-algebra)} & \makecell{Deformed \\ commutators} & \makecell{Spin-network \\ discreteness} & \makecell{Entanglement \\ wedges} \\
\hline
Entropy correction & \(1/M^4\) & \(1/M^2\), (log)~\cite{Medved2004}
 & (log), \(1/A\)~\cite{AshtekarBojowald2005} & \makecell{Non-perturbative \\ area change} \\
\hline
QNM shift scaling & \(1/M^6\) & \(1/M^2\)--\(1/M^4\)~\cite{Scardigli1999} & \(1/M^2\), (log)~\cite{Barrau2014} & \makecell{Model-dependent \\ (via echoes)} \\
\hline
Hawking temperature & Lowered (geometry) & Modified spectrum & Typically lowered & Indirect \\
\hline
Echoes & From regular core~\cite{Abedi2017,Conklin2018} & Possible & Rarely addressed & From late islands~\cite{Raju2021Lessons,Abedi2017} \\
\hline
Hilbert space & \makecell{Non-factorizable: \\ \(\mathcal{H}_+ \oplus \mathcal{H}_-\)} & Standard & \makecell{Spin network \\ quantization} & \makecell{Semiclassical with \\ auxiliary interior} \\
\hline
Covariant? & Yes & Often no & Yes & Yes (in AdS/CFT) \\
\hline
Needs holography? & No & No & No & Yes \\
\hline
Evaporation behavior & \makecell{Unitarity-preserving;\\ core smoothed by geometry} & \makecell{Non-unitary or cutoff\\ dependent} & \makecell{Unitary; singularity\\
  resolved via quantum geometry~\cite{AshtekarBojowald2005}}
& \makecell{Unitary; late-time\\ island correlations~\cite{Raju2021Lessons}}\\
%\makecell{Unitary; late-time\\ island correlations} \\
\hline
Horizon structure & \makecell{Modified but present} & \makecell{Modified trans-Planckian \\ region} & \makecell{Preserved with quantum \\ corrections} & \makecell{Absent; replaced by \\ wedge / reflective surface~\cite{Barcelo2009}}\\
\hline
\end{tabular}
}
\caption*{\scriptsize {Note:} “(log)” refers to logarithmic
  corrections—either to black hole entropy (e.g., \( S \sim A/4 -
  \alpha \log A \)) or to QNM frequency
  shifts. Such corrections typically arise as subleading quantum
  gravity effects in theories like LQG, GUP, and string theory.}
\label{tab:comparison}
\end{table}

\section{Observational Signatures}\label{sec:observables}

While pseudo-complex gravity is fundamentally a theoretical framework,
its modifications to spacetime structure at high curvature suggest
potential observational consequences—especially in the strong gravity
regimes near black hole horizons and in the early universe. In this
section, we outline several avenues where pc-Gravity may lead to
testable deviations from general relativity.

\subsection{Black Hole Shadows and Imaging}

The addition of a pseudo-complex correction to the Schwarzschild
metric (Eq.~(\ref{eq:ds2Schw})) modifies the effective potential experienced by photons
orbiting a black hole. Specifically, the shift in the photon sphere
and the behavior of light rays near the horizon can lead to measurable
differences in the shape and size of black hole shadows.

Observations from the Event Horizon Telescope (EHT), which has imaged
the shadows of M87* and Sagittarius A*, are placing increasingly
stringent constraints on deviations from the classical Kerr
geometry. In pseudo-complex gravity (pc-Gravity), the correction term
\( B/(6 r^4) \) in the metric modifies light bending near the black
hole, potentially shifting the angular size of the shadow or altering
the structure of the photon ring. Future very-long-baseline
interferometric arrays could be sensitive to such deviations,
especially for supermassive black holes whose mass and spin are
independently constrained.

These effects may help distinguish pc-Gravity from horizonless models
such as black stars~\cite{Barcelo2009}, which also predict
modifications to shadow structure. A key distinction is that
pc-Gravity preserves a continuous, though modified, horizon, while
black star models eliminate the horizon altogether. This difference
could manifest in specific features of the photon ring or in the
polarization pattern near the shadow edge, offering a potential
observational handle.

Although some of the predicted signatures overlap with those of more
general exotic compact objects (ECOs), the presence of a smoothed but
intact horizon in pc-Gravity may lead to observable differences in the
ringdown waveform, the timing and structure of gravitational wave
echoes, or subtle deviations in shadow
substructure~\cite{Cardoso2016}.

\subsection{Gravitational Wave Ringdowns}

The late-time behavior of gravitational waves emitted from black hole
mergers—known as the ringdown phase, is determined by the quasi-normal
mode spectrum of the remnant object. In pc-Gravity, the modified
horizon structure and regularized interior lead to changes in the
boundary conditions that determine the QNM frequencies.

Detectable deviations from the expected general relativistic QNM
spectrum could signal the presence of new physics at the horizon
scale. Observations from LIGO, Virgo, and KAGRA, and future
space-based detectors such as LISA, may be sensitive to such
corrections. Moreover, the absence of a singularity in pc-Gravity
could give rise to new late-time echo signals or deviations in the
damping rates of QNMs, potentially matching the phenomenology of
echo templates like those in~\cite{Conklin2018}.

\subsection{Primordial Signatures in the Early Universe}

If pc-Gravity is applicable at Planckian energies, it may also
influence the dynamics of the early universe, particularly during the
inflationary epoch. The presence of a maximal acceleration or minimal
length scale could suppress high-frequency modes in the primordial
power spectrum, leading to observable effects in the cosmic microwave
background (CMB).

In particular, the scalar and tensor spectra may exhibit
scale-dependent features, such as running of the spectral index or
suppression of small-scale power, that differ from standard
inflationary predictions. These effects could be probed by future CMB
polarization experiments and large-scale structure surveys.

\subsection{Deviations in Compact Object Structure}

Pseudo-complex gravity may also alter the structure of neutron stars
and other compact objects by modifying the Tolman–Oppenheimer–Volkoff
 equation \cite{Hess2016}. This could lead to changes in the
maximum mass, radius, and tidal deformability of such
objects. Observational constraints from pulsar timing, X-ray
measurements (e.g., NICER), and gravitational wave detections from
neutron star mergers (e.g., GW170817) can thus be used to place limits
on the scale of pseudo-complex corrections.

\subsection{Prospects and Challenges}

While the effects of pc-Gravity are expected to be small in weak-field
regimes, its predictions in high-curvature settings are distinctive
and potentially observable. A major challenge lies in developing
detailed models of astrophysical observables within this framework and
comparing them to current and future data. Nonetheless, the covariant
and geometric nature of the theory makes it well suited for coupling
with observational signatures, offering a promising route toward
empirical validation or falsification. In particular, echo signal
templates developed in~\cite{Conklin2018} could be adapted to test the
structure predicted by pc-Gravity.

\subsection{Implications for Black Hole Mergers and Ringdown Observables}

Black hole mergers provide a unique laboratory for probing
strong-field gravity, especially during the ringdown phase where the
emitted gravitational wave signal is governed by the quasi-normal mode
 spectrum of the final remnant. Pseudo-complex gravity, by
modifying the near-horizon and interior structure of black holes, can
leave distinctive signatures in this regime.

\paragraph*{Frequency Shifts in Quasi-Normal Modes.} 
The pseudo-complex correction \( B/(6 r^4) \) alters the effective
potential that governs linear perturbations of the Schwarzschild
background. As shown in Section~VII, this leads to a shift in both the
real and imaginary parts of the QNM frequencies. The fractional shift
for the fundamental mode can be approximated as
\begin{equation}
\frac{\delta \omega}{\omega} \sim -\frac{B c^4}{6 G^2 M^6}, \quad
\rm{with} \quad B = \ell^4 ,
\label{eq:qnm_shift}
\end{equation}
where \( \ell \) is the pseudo-complex minimal length scale and \( r_s
= 2GM/c^2 \) is the classical Schwarzschild radius.

This scaling becomes negligible for astrophysical black holes but
grows significant near the Planck mass, where quantum gravitational
effects are expected to dominate. While the correction is
suppressed for stellar-mass black holes with \( \ell \sim \ell_P \), it
could become detectable if \( \ell \) is effectively larger, or for
lower-mass black holes such as those possibly formed in the early
universe.

\begin{figure}[ht]
    \centering
    \includegraphics[width=0.65\textwidth]{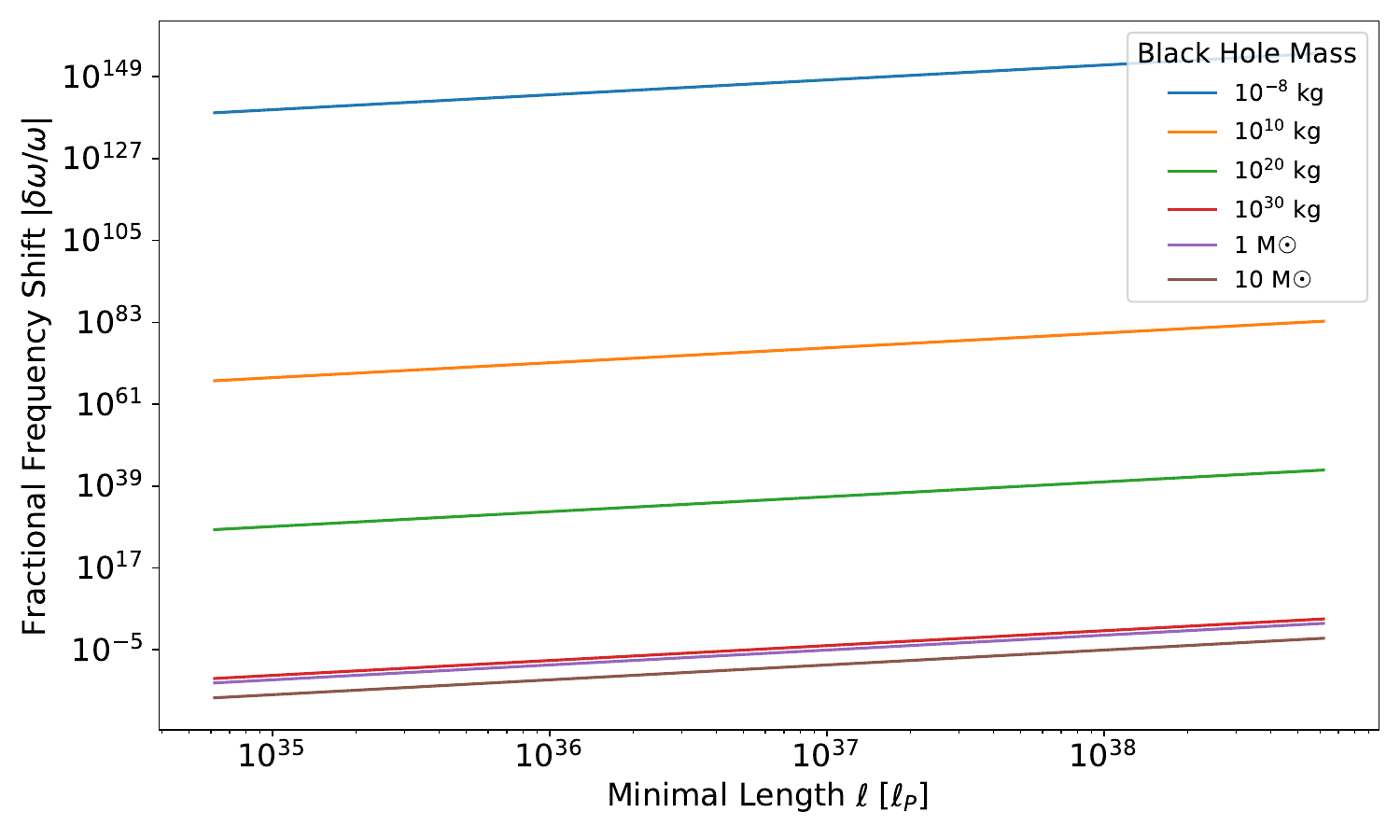}
    \caption{Fractional QNM frequency shift \(
      |\delta \omega / \omega| \) as a function of the minimal length
      scale \( \ell \), expressed in Planck units \( \ell_P \), for a
      range of black hole masses. The correction scales as \( (\ell^2
      / r_s^2)^2 \), leading to negligible deviations for
      astrophysical black holes at Planckian \( \ell \), but
      potentially observable effects for primordial or
      near-Planck-scale black holes, or for larger values of \( \ell \).
      This parametric scan illustrates how gravitational wave
      observations may constrain or detect ultraviolet modifications
      in pseudo-complex gravity.}
    \label{fig:qnm-shift-vs-ell}
\end{figure}

To quantify the observational relevance of these effects, we perform a
parametric scan over the minimal length scale \( \ell \), treating it
as a free parameter above the Planck length.
Figure~\ref{fig:qnm-shift-vs-ell} shows the resulting
fractional QNM frequency shift \( |\delta \omega / \omega| \) as a
function of \( \ell \) for several black hole masses. The corrections
scale as \( (\ell^2 / r_s^2)^2 \), and become appreciable only when \(
\ell \) exceeds \( 10^3 \, \ell_P \) for stellar-mass black holes. In
contrast, for low-mass or primordial black holes, even Planck-scale
values of \( \ell \) yield order-unity effects. This analysis
underscores the potential of gravitational wave observations to probe
ultraviolet structure in pseudo-complex gravity and motivates future
bounds on \( \ell \) from the absence of observed deviations from general
relativity.

\paragraph*{Late-Time Echoes.} If the classical singularity is replaced by a
regular core, as predicted in pc-Gravity, perturbations may reflect
off the core and generate late-time echoes in the gravitational wave
signal \cite{Cardoso2016}. These echoes would appear as repeated
bursts following the main ringdown, with characteristic time delays
set by the round-trip travel time between the effective potential peak
and the core. Phenomenological models capturing this echo
structure—based on transfer functions and signal windows—have been
developed in~\cite{Conklin2018}.  Although current searches have not
found conclusive evidence for such signals, future detectors with
higher sensitivity and stacking techniques across events may reveal
these features. Initial analyses of LIGO data suggested tentative
evidence for such echo signals following black hole
mergers~\cite{Abedi2017}, prompting further searches across multiple
observational runs.

To assess the observational viability of such signatures, it is
instructive to treat the minimal length scale \( \ell \) as a free
parameter and explore its effect on measurable quantities such as
quasi-normal mode frequencies and Hawking temperature. While
theoretical consistency motivates setting \( \ell \sim \ell_P \),
modest deviations above the Planck scale are not excluded by first
principles and may yield testable signatures. For example, increasing
\( \ell \) by an order of magnitude amplifies QNM frequency shifts by
a factor of \( 10^4 \), due to the \( \ell^4 \) scaling in the
correction term \( B = \ell^4 \). This significantly enhances the
prospects for detection in near-horizon physics. Using Eq.~(27), one
can estimate that for stellar-mass black holes, deviations become
marginally detectable for \( \ell \gtrsim 10^3 \, \ell_P \),
corresponding to \( \delta \omega/\omega \sim 10^{-64} \), still well
below current observational precision but within potential reach of
third-generation detectors. Conversely, for low-mass primordial black
holes, even Planck-scale values of \( \ell \) can produce \(
\mathcal{O}(1) \) effects. This parametric scan underscores the value
of gravitational wave observations as a probe of ultraviolet
modifications to gravity and allows bounds to be placed on \( \ell \)
in the absence of observed deviations from general relativity.

\paragraph*{Modified Evaporation and Remnant Scenarios.} In the merger of
near-Planckian mass black holes or hypothetical primordial black
holes, pseudo-complex corrections could suppress the Hawking
temperature and modify the endpoint of evaporation. This raises the
possibility of stable remnants or long-lived metastable cores, which
may have observable consequences in gravitational wave or high-energy
astrophysical contexts.

With ongoing and upcoming gravitational wave observations
from LIGO to Virgo to KAGRA, and future missions such as LISA and the
Einstein Telescope, precise measurements of QNM spectra and late-time
signal structure could provide empirical tests of pseudo-complex
gravity. In particular, constraints on deviations in QNM frequencies
at the sub-percent level, or the detection of post-merger echoes,
would serve as powerful probes of UV-complete gravity models such as
the one explored here.

\section{Quasi-Normal Mode Shifts and Quantum Gravity Comparisons}\label{sec:qnm-comparison}

Building on the estimate in Eq.~(\ref{eq:qnm_shift}), the presence of
a pseudo-complex correction term \( B/(6r^4) \) in the Schwarzschild
metric modifies the effective potential governing gravitational
perturbations around a black hole.  To estimate the impact on
quasi-normal mode frequencies, we consider how this correction alters
the location and height of the peak of the Regge–Wheeler potential.

For scalar or gravitational perturbations in a Schwarzschild
background, the leading-order QNM frequency is given approximately by
the WKB expression~\cite{schutzwill1985}
\begin{equation}
\omega \sim \Omega_c - i \Lambda,
\end{equation}
where \( \Omega_c \) is the orbital frequency at the photon sphere \(
r = r_c \) and \( \Lambda \) is the Lyapunov exponent that controls
damping.

In standard GR, the photon sphere lies at \( r_c = 3GM/c^2 \).  To
evaluate the shift in the photon sphere, consider the modified metric
in Eq.~(\ref{eq:g_tt.2}), which incorporates the pseudo-complex
correction
\begin{equation}
g_{tt}(r) = 1 - \frac{2GM}{c^2 r} + \frac{B}{6 r^4},
\label{eq:g_tt.2}
\end{equation}
Solving for the extremum of the effective potential (via null geodesics), one finds to first order in \( B \)
\begin{equation}
r_c^{\rm{pc}} \approx \frac{3GM}{c^2} \left(1 - \frac{4 B c^4}{27 G^2 M^2} \right).
\label{eq:rc_pc}
\end{equation}
The orbital frequency is related to \( r_c \) by
\begin{equation}
\Omega_c = \frac{c^3}{\sqrt{27} GM},
\end{equation}
so the fractional shift in frequency is:
\begin{equation}
\frac{\delta \omega}{\omega} \sim \frac{\delta r_c}{r_c} \sim -\frac{4
  B c^4}{27 G^2 M^2}.
  \label{eq:qnm_shift.2}
\end{equation}

Setting \( B = \ell^4 \) with \( \ell \sim \ell_P \), we estimate the
fractional shift for a solar-mass black hole (\( M \sim 2 \times
10^{30} \,\mathrm{kg} \))
\begin{equation}
\frac{\delta \omega}{\omega} \sim -\frac{4 \ell^4 c^4}{27 G^2 M^4}
= -\frac{4Bc^4}{27 G^2 M^4}
= -\frac{4}{27} \left( \frac{\ell^2}{r_s^2} \right)^2
\approx -10^{-76}.
\label{eq:delta_omega_final}
\end{equation}

This is well below current experimental precision. However, for
hypothetical Planck-scale black holes (\( M \sim m_P \)), the
correction becomes \( \mathcal{O}(1) \), implying that QNM spectra
could be dramatically altered in the high-curvature regime. For
intermediate masses (e.g., primordial black holes), deviations could
be detectable with future ultra-sensitive detectors such as LISA or
advanced third-generation ground-based observatories.  Future
third-generation detectors may be able to constrain pseudo-complex
length scales as small as \( \ell \sim 10^{-20} \, \rm{m} \), still
many orders of magnitude above the Planck length, thereby opening a
potential observational window onto pc-Gravity in the high-curvature
regime.

This order-of-magnitude estimate highlights a key feature of
pc-Gravity that corrections are negligible at astrophysical scales but
become dominant near the Planck regime. The structure of QNM spectra
may thus serve as a probe of pseudo-complex corrections in
hypothetical early-universe or micro black holes.

To contextualize the pseudo-complex gravity prediction for QNM
frequency shifts, we compare its scaling behavior with those arising
in other quantum gravity frameworks. Generalized uncertainty principle
 models typically predict a frequency suppression of the form
\begin{equation}
\frac{\delta \omega}{\omega} \sim -\beta \left( \frac{m_P}{M} \right)^2,
\end{equation}
where \( \beta \sim \mathcal{O}(1) \) is a model-dependent constant
and \( m_P \) is the Planck mass~\cite{Kempf1995}. Similarly, loop
quantum gravity (LQG) and polymer quantization schemes suggest entropy
and area quantization effects that lead to QNM corrections of the
form
\begin{equation}
\frac{\delta \omega}{\omega} \sim -\frac{\alpha}{M^2},
\end{equation}
where \( \alpha \) encodes the strength of quantum geometric corrections.

\begin{figure}[ht]
    \centering
    \includegraphics[width=0.6\linewidth]{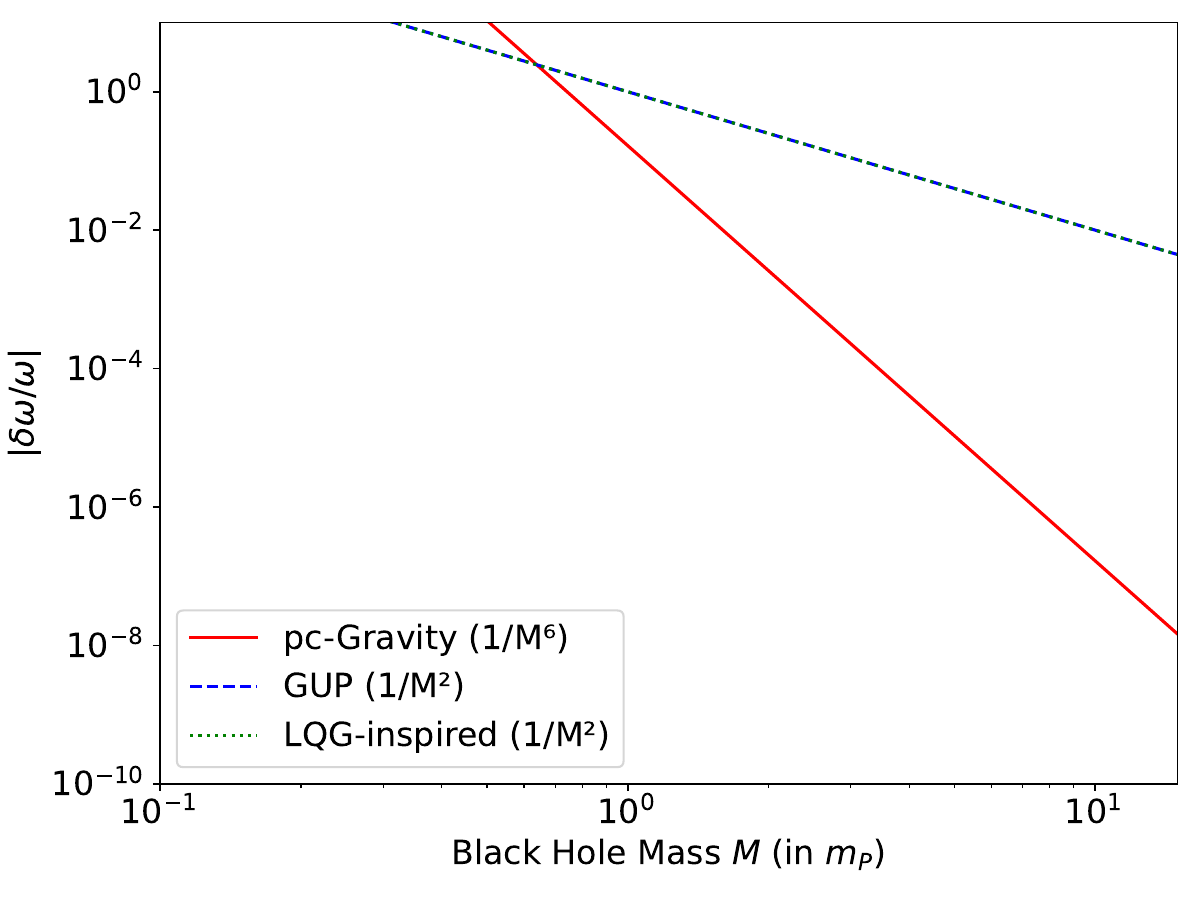}
    \caption{Comparison of fractional shifts in the QNM frequency \(
      \delta \omega / \omega \) as predicted by three models:
      pseudo-complex gravity, generalized uncertainty principle (GUP),
      and loop quantum gravity (LQG)-inspired corrections. Each model
      yields negligible deviations at astrophysical mass scales, but
      diverges significantly near the Planck scale. The pc-Gravity
      curve (red) reflects \( 1/M^6 \) scaling from a \( B/(6r^4) \)
      correction that becomes dominant for \( M \sim m_P \), similar
      in magnitude to GUP (blue dashed) and LQG (green dotted)
      corrections. This comparison situates pc-Gravity within the
      broader quantum gravity landscape and highlights its unique
      geometric foundation.  }
    \label{fig:qnm_shift_comparison}
\end{figure}

Figure~\ref{fig:qnm_shift_comparison} shows these three
models, pseudo-complex gravity, GUP, and LQG-inspired, using natural
values for \( \beta \) and \( \alpha \). All models agree in
predicting negligible corrections for astrophysical black holes, but
diverge significantly in the Planckian regime. Notably, pc-Gravity
provides a geometric, covariant route to UV suppression without
modifying the commutator algebra or assuming spacetime discreteness.

\section{Conclusion and Outlook}\label{sec:conclusion}

In this work, we have explored the implications of pseudo-complex
gravity for the black hole information paradox. By extending spacetime
to include pseudo-complex coordinates, pc-Gravity introduces a
covariant ultraviolet (UV) regulator that smooths classical
singularities and modifies the geometry near black hole horizons. This
geometric extension yields several significant outcomes relevant to
quantum gravity.

We showed that pc-Gravity modifies the Schwarzschild metric by
introducing a correction term of the form \( B/(6 r^4) \), which
regularizes the curvature at \( r = 0 \) and alters the location of
the event horizon.  This regularization is quantified by the softened
behavior of the Kretschmann scalar in Eq.~(\ref{eq:K_pc}), where the
leading divergence is suppressed by the pseudo-complex structure.  The
modifications result in finite entropy and a corrected
Bekenstein--Hawking formula with subleading terms that scale as \(
1/M^2 \), as given in Eq.~(\ref{eq:S_pc.b}). This behavior is
consistent with predictions from generalized uncertainty principles,
non-commutative geometry, and loop quantum gravity, but arises here
from a smooth, continuous deformation of spacetime geometry.

Importantly, we argued that pc-Gravity may soften or resolve the black
hole information paradox by modifying the causal and entanglement
structure of spacetime. The pseudo-complex regularization permits
deviations from thermality in Hawking radiation, allows information to
be retained or transferred across a regularized core, and supports
unitarity-preserving evaporation processes—though a complete dynamical
model of this process remains an open direction for future work.
Conceptually similar goals are pursued in models with a closed
apparent horizon~\cite{Frolov2014Nonsingular}, which maintain regular
geometry and causal continuity throughout the evaporation process.
This is reminiscent of the unitary evaporation paradigm in loop
quantum gravity, where the singularity is resolved and no information
is lost~\cite{AshtekarBojowald2005}.

We also compared pseudo-complex gravity with several leading
approaches to the black hole information paradox, including firewalls,
fuzzballs, holographic islands, and string-theoretic models. While
these frameworks share certain foundational aims, pc-Gravity
distinguishes itself by incorporating ultraviolet regularization
directly into the spacetime manifold through a covariant geometric
extension, without invoking non-locality, symmetry breaking, or
discrete structures.

Beyond its conceptual contributions, pc-Gravity yields observational
signatures. Modifications to black hole shadows, quasi-normal mode
 spectra, and early-universe perturbations offer potential
empirical access to the theory. Upcoming data from the Event Horizon
Telescope, LIGO/Virgo, LISA, and cosmic microwave background (CMB)
measurements may provide meaningful constraints on pseudo-complex
corrections.

Tentative evidence for post-merger gravitational wave
echoes~\cite{Abedi2017} further highlights the relevance of models
like pc-Gravity, which predict reflective core structures and
late-time deviations from classical ringdown behavior.

In addition to its theoretical appeal, pseudo-complex gravity yields
potentially testable predictions in the context of black hole
mergers. The modification of the Schwarzschild metric introduces
subleading corrections to the QNM spectrum, which may manifest as
small frequency shifts during the ringdown phase of gravitational wave
signals. We estimate the magnitude of these corrections and compare
them with those predicted by generalized uncertainty principles and
loop quantum gravity. While the deviations are negligible for
stellar-mass black holes assuming a Planck-scale minimal length, they
become significant for near-Planckian or primordial black holes and
could be probed by next-generation gravitational wave
observatories. Furthermore, the regularized core structure predicted
by pseudo-complex gravity may give rise to late-time echoes or remnant
signatures, offering new observational windows into
ultraviolet-complete gravity in the strong-field regime.
While these signatures overlap with those predicted by horizonless ECO
models, pc-Gravity retains a smoothed horizon rather than replacing it
altogether, which could yield distinct signatures in the ringdown
phase, echo spacing, or shadow substructure.

Looking ahead, several promising directions merit further
exploration. Chief among them is the development of a fully quantum or
semiclassical model of black hole evaporation within the
pseudo-complex framework. Additionally, investigating entanglement
wedge reconstruction and modular flow in pseudo-complex geometries
could provide new insight into quantum information dynamics in
gravitational systems. Extending the theory to include rotating and
charged black holes is another important step, as such solutions may
exhibit richer causal structures, nontrivial ergospheres, and modified
thermodynamic behavior beyond the Schwarzschild case.

A particularly important extension involves generalizing
pseudo-complex gravity to axisymmetric spacetimes, most notably the
Kerr solution. In classical general relativity, the Kerr metric
describes a rotating black hole with distinct features such as an
ergosphere, frame dragging, and an inner Cauchy horizon. These
structures introduce additional instabilities and raise further
questions about information retention and determinism. Within the
pc-Gravity framework, one anticipates that the ring-like singularity
of the Kerr solution would be regularized, just as the central
singularity is smoothed in the Schwarzschild case. Moreover, the
pseudo-complex extension may alter the structure of the ergosphere and
shift the locations of the horizons, potentially affecting energy
extraction processes such as the Penrose mechanism and
superradiance. Although a full pseudo-complex generalization of the
Kerr metric remains to be derived, a perturbative or slow-rotation
expansion could offer a viable approach. This would provide further
insight into how pseudo-complex geometry shapes the causal and
thermodynamic properties of rotating black holes.

Finally, connecting the pseudo-complex algebra to path integral or
spin foam formulations of quantum gravity may help embed the theory
within the broader landscape of nonperturbative quantum gravity
approaches. 

Taken together, these results position pseudo-complex gravity as a
novel and promising framework for addressing some of the deepest
challenges in quantum gravity. Its intrinsic geometric regularization
mechanism, compatibility with classical general relativity, and
potential for observational verification underscore its value both as
a standalone theory and as a bridge to more established quantum
gravity paradigms. Continued development of the pseudo-complex
formalism, particularly its application to rotating spacetimes, quantum
field dynamics, and gravitational phenomenology, may further clarify
its role as a viable extension of general relativity in the
ultraviolet regime.

%%%%%%%%%%%%%%%%%%%%%%%%%%%%%%%%%%%%%%%%%%%%%%%%%%%%%%%%%%%%%%%%%%%%%%%

\appendix

\section{Scalar Green's Function and Regulated Two-Point Function
  in pc-AdS}\label{sec:appendix-a}
\setcounter{equation}{0}
\renewcommand{\theequation}{A\arabic{equation}}

In this appendix, we derive the regulated boundary two-point function
for a scalar field propagating in a pseudo-complex AdS (pc-AdS)
spacetime. The pseudo-complex formalism introduces an internal
Gaussian smearing over the pseudo-imaginary coordinate \( y^\mu \),
implementing a covariant UV cutoff.
 
\subsection{Bulk Scalar Action and Field Decomposition}

We begin with the action for a massive scalar field \( \Phi(X) \) in a
pc-AdS\(_{d+1} \) background,
\begin{equation}
S = \frac{1}{2} \int d^{d+1}X \, \sqrt{-G} \left( G^{MN} \partial_M
\Phi \partial_N \Phi + M^2 \Phi^2 \right),
\end{equation}
where \( X^M = (x^\mu, z, y^\mu) \) are pseudo-complex coordinates and
\( G_{MN} \) is the pseudo-complex bulk metric. We decompose the field
as
\begin{equation}
\Phi(X) = \phi(x, z) + I \varphi(x, z),
\end{equation}
where both \( \phi \) and \( \varphi \) are real-valued fields in AdS.

\subsection{Smearing over Pseudo-Imaginary Directions}

In the pc-AdS framework, the pseudo-imaginary directions \( y^\mu \)
are not physical coordinates but internal degrees of freedom
associated with regularization. The scalar field is extended as
\begin{equation}
\Phi(z, x, y) = \int d^d x' \, d^d y' \, K(z, x; x') \, f_\ell(y - y') \, \phi_0(x'),
\end{equation}
where
   \( \phi_0(x') \) is the boundary source field,
   \( K(z, x; x') \) is the standard AdS bulk-to-boundary propagator, and
  \( f_\ell(y - y') \) is a Gaussian regulator with width \( \ell \).
The Gaussian kernel is defined as
\begin{equation}
f_\ell(y) = \left( \frac{1}{\pi \ell^2} \right)^{d/2}
\exp\left(-\frac{y^2}{\ell^2}\right).
\end{equation}

\subsection{Boundary Two-Point Function}

The boundary two-point function is obtained by integrating over the
internal \( y^\mu \) direction,
\begin{eqnarray}
\langle \mathcal{O}(x) \mathcal{O}(x') \rangle_{\rm{pc}} &=& \int d^d
y \, f_\ell(y) \, \frac{1}{(|x - x'|^2 + y^2)^\Delta}  \\ &=&
\frac{1}{(|x - x'|^2 + \ell^2)^\Delta}.
\end{eqnarray}

The result follows from the identity
\begin{equation}
\int d^d y \left( \frac{1}{\pi \ell^2} \right)^{d/2}
\frac{e^{-y^2/\ell^2}}{(|x - x'|^2 + y^2)^\Delta} = \frac{1}{(|x -
  x'|^2 + \ell^2)^\Delta},
\end{equation}
which can be derived by rescaling and using spherical coordinates in \( y \)-space.

\subsection{Interpretation and Significance}

This result shows that the pseudo-complex extension introduces a
nonperturbative regulator into the boundary correlators. As \( \ell
\to 0 \), we recover the standard CFT result
\[
\langle \mathcal{O}(x) \mathcal{O}(x') \rangle \sim \frac{1}{|x - x'|^{2\Delta}}.
\]
For finite \( \ell \), the correlator remains finite as \( x \to x'
\), thereby avoiding the UV divergence and rendering the operator
product expansion well-defined at all distances.

This mechanism is reminiscent of similar effects in non-commutative
field theories and generalized uncertainty principles, but here it
arises geometrically from the extended pseudo-complex manifold and not
from modifying the algebra of fields or operators.

\section{Modified Surface Gravity and Hawking Temperature in pc-Gravity}\label{sec:appendix-b}
\setcounter{equation}{0}
\renewcommand{\theequation}{B\arabic{equation}}

In this appendix, we compute the surface gravity and Hawking
temperature for the pseudo-complex corrected Schwarzschild
geometry. This analysis reveals how the built-in UV regularization of
pc-Gravity modifies the thermodynamics of black holes and suggests
potential observational implications.

\subsection{Metric and Horizon Location}

We begin with the modified Schwarzschild metric incorporating the
pseudo-complex correction
\begin{eqnarray}
  ds^2 &=& \left( 1 - \frac{2GM}{c^2 r} + \frac{B}{6r^4} \right) c^2 dt^2 \\
  && - \left( 1 - \frac{2GM}{c^2 r} + \frac{B}{6r^4} \right)^{-1} dr^2 - r^2 d\Omega^2,
\end{eqnarray}
where \( B \propto \ell^4 \) introduces a geometric cutoff scale.

The event horizon \( r_+ \) is determined by solving \( g_{tt}(r_+) = 0 \):
\begin{equation}
1 - \frac{2GM}{c^2 r_+} + \frac{B}{6 r_+^4} = 0.
\end{equation}

Solving to leading order yields
\begin{equation}
r_+ \approx \frac{2GM}{c^2} - \frac{B c^2}{24 G^2 M^3}.
\end{equation}

\subsection{Surface Gravity}

The surface gravity \( \kappa \) is defined via
\begin{equation}
\kappa = \frac{1}{2} \left| \frac{d g_{tt}}{dr} \right|_{r = r_+}.
\end{equation}

From the metric
\begin{equation}
g_{tt}(r) = 1 - \frac{2GM}{c^2 r} + \frac{B}{6r^4},
\end{equation}
so the derivative becomes
\begin{equation}
\frac{d g_{tt}}{dr} = \frac{2GM}{c^2 r^2} - \frac{2B}{3 r^5}.
\end{equation}

Evaluating this at \( r = r_+ \) gives
\begin{equation}
\kappa = \frac{GM}{c^2 r_+^2} - \frac{B}{3 r_+^5}.
\end{equation}

Expanding to leading order in \( B \), using \( r_+ \approx 2GM/c^2 \), we find
\begin{equation}
\kappa \approx \frac{c^4}{4GM} \left( 1 - \frac{5 B c^4}{24 G^2 M^4} \right).
\end{equation}

\medskip

\subsection{Corrected Hawking Temperature}

The Hawking temperature is related to the surface gravity via
\begin{equation}
T_H = \frac{\hbar \kappa}{2\pi k_B}.
\end{equation}

Substituting the corrected expression for \( \kappa \), we obtain
\begin{equation}
T_H^{\rm{pc}} \approx \frac{\hbar c^3}{8\pi G M k_B} \left( 1 -
\frac{5 B c^4}{24 G^2 M^4} \right).
\end{equation}

This shows that the Hawking temperature is lowered by the presence of
the pseudo-complex regulator. For large black holes, the correction is
negligible, but it becomes significant near the Planck scale.

The negative correction to the temperature implies a slower
evaporation rate for small black holes. This is consistent with the
idea that a minimal length prevents the formation of arbitrarily
high-energy curvature regions. Additionally, the entropy and
temperature corrections together suggest the possibility of a stable
remnant or modified late-time evaporation behavior, a key ingredient
in potential resolutions of the black hole information paradox.

\section*{Appendix C. Sketch of a Quantum Field Quantization Strategy in
  pc-Gravity}\label{sec:appendix-c}
\setcounter{equation}{0}
\renewcommand{\theequation}{C\arabic{equation}}

To assess the prospects for a full quantum formulation of
pseudo-complex gravity, we sketch a preliminary strategy for the
quantization of a free scalar field on a pseudo-complex (pc)
manifold. The goal is to explore how standard quantum field theory
(QFT) structures, such as Fock space, commutators, and correlation
functions, may be extended to accommodate the geometric regularization
built into pc-Gravity.

\subsection*{C.1. Field Decomposition and Mode Expansion}

Let $X^\mu = x^\mu + I y^\mu$ denote a point in the
pseudo-complexified spacetime, and consider a real pseudo-complex
scalar field $\Phi(X)$ with the decomposition
\begin{equation}
\Phi(X) = \phi(x) + I \psi(x),
\end{equation}
where $\phi$ and $\psi$ are real-valued fields defined on the base
manifold. One can promote both $\phi(x)$ and $\psi(x)$ to
operator-valued distributions
\begin{equation}
\hat{\Phi}(X) = \hat{\phi}(x) + I \hat{\psi}(x).
\end{equation}
A natural choice is to expand these in terms of a complete set of mode functions
\begin{eqnarray}
  \hat{\phi}(x) &=& \int \frac{d^3 \vec{k}}{(2\pi)^3
    2\omega_{\vec{k}}} \left( a_{\vec{k}} u_{\vec{k}}(x) +
  a_{\vec{k}}^\dagger u_{\vec{k}}^*(x) \right), \\
\hat{\psi}(x) &=& \int \frac{d^3 \vec{k}}{(2\pi)^3 2\omega_{\vec{k}}}
\left( b_{\vec{k}} v_{\vec{k}}(x) + b_{\vec{k}}^\dagger
v_{\vec{k}}^*(x) \right),
\end{eqnarray}
where $\omega_{\vec{k}} = \sqrt{\vec{k}^2 + m^2}$ and the ladder
operators $a_{\vec{k}}, b_{\vec{k}}$ obey canonical commutation
relations.

\subsection*{C.2. Commutation Relations and Fock Space}

Assuming bosonic statistics, the operators satisfy
\begin{eqnarray}
  \left[ a_{\vec{k}}, a_{\vec{k}'}^\dagger \right] &=& (2\pi)^3 \delta^{(3)}(\vec{k} - \vec{k}') \\
\left[ b_{\vec{k}}, b_{\vec{k}'}^\dagger \right] &=& (2\pi)^3 \delta^{(3)}(\vec{k} - \vec{k}') \\
\left[ a_{\vec{k}}, b_{\vec{k}'} \right] &=& \left[ a_{\vec{k}}, b_{\vec{k}'}^\dagger \right] = 0
\end{eqnarray}
The vacuum state $| 0 \rangle$ is annihilated by both
$a_{\vec{k}}$ and $b_{\vec{k}}$, and the Fock space is constructed by
acting with their respective creation operators.

\subsection*{C.3. Two-Point Function and Regulator}

The pc-structure leads to a modified Wightman function. Let
$f_\ell(y)$ be a Gaussian regulator over the pseudo-imaginary
components
\begin{equation}
f_\ell(y) = \left(\frac{1}{\pi \ell^2}\right)^{d/2} \exp\left(-\frac{y^2}{\ell^2}\right).
\end{equation}
The regulated two-point function is defined as
\begin{eqnarray}
\langle 0 | \hat{\Phi}(X) \hat{\Phi}(X') | 0 \rangle &=& \int d^d y \,
f_\ell(y) \, \langle 0 | \hat{\phi}(x) \hat{\phi}(x') | 0 \rangle \\
&=&
\frac{1}{\left(|x - x'|^2 + \ell^2\right)^\Delta},
\end{eqnarray}
in agreement with Appendix A. This suggests that the UV divergence of
the coincidence limit $x \to x'$ is suppressed by the pc-regulator
$\ell$.

%\subsection*{C.4. Interpretation}

This framework hints at a doubled field content $(\phi, \psi)$ and a
natural smearing over internal degrees of freedom. The algebraic
sector structure of pseudo-complex numbers (e.g., via idempotents
$e_\pm$) may give rise to braided or non-factorizable Fock sectors,
potentially altering entanglement structure. While this construction
is schematic, it establishes a platform for path-integral, spin-foam,
or algebraic QFT extensions in pc-Gravity.

\section*{Acknowledgment} POH acknowledges financial support from DGAPA-PAPIIT (IN116824).

\bigskip

\bibliographystyle{iopart-num}
\bibliography{pc-GR.bib}

\end{document}